In the memory of H.Hertz.

# The new sight on the Hertz's electrodynamics of a moving bodies.

Pechenkov A.N.
Institute of Metal Physics, Ural Branch of the Rus. Acad. Sci., Ekaterinburg.
pechenkov@imp.uran.ru

A little error was eliminated from Hertz's equations. New Hertz's equations don't contradict to all electromagnetic experiments. Therefore Hertz's electrodynamics is the alternative to Einstein's electrodynamics. It means that the question about the absolute or relative character of both space and time must be rediscussed. Lorentz's transformations can be used only if our theoretical model of an electromagnetic phenomenon is not complete one .

Hertz's equations.

During more than 110 years the Hertz's equations for the electrodynamics of a moving bodies [1] are referred in scientific literature as an example of an unsuccessful attempt to create the electrodynamics of a moving bodies in absolute time and in absolute space. The careful analysis of the critical works ([2 – 4], for example) shows us that the authors right found an error in Hertz's equations. But they didn't try to eliminate the error from the equations. Meanwhile it is very simple error. The error consists in equality of both charge velocity and ether velocity ("full ether drugging" by any charges). The error leads to contradictions of Hertz's equations with a number of electromagnetic experiments (for example, Roentgen W., Eichenwald A., Wilson H. experiments, see below).

At the recent time we should not be frightened the idea of an ether. This idea is a basis of contemporary physics. Still Einstein spoke that the physics needs in an ether. But it must be, of course, not mechanical ether. In the work I want to show that the Hertz's equations without the error can be very flexible for the explanation of all electromagnetic experiments. We must to replace only one letter in the Hertz's equations.

Hertz died 37 years old, only a few years after publication of he's equations [1]. It is very probably that Hertz could correct the little error in he's equations if he's life would be more long.

So, the original Hertz's equations are [1-4]:

$$rot(\mathbf{H}) = \frac{\partial \mathbf{D}}{\partial t} + rot[\mathbf{D}x\mathbf{u}] + \rho\mathbf{u} + \mathbf{j}; \quad (1)$$

$$rot(\mathbf{E}) = -\frac{\partial \mathbf{B}}{\partial t} + rot[\mathbf{B}x\mathbf{u}]; \quad (2)$$

$$div(\mathbf{B}) = 0; \quad (3)$$

$$div(\mathbf{D}) = \rho; \quad (4)$$

here: $\rho$ – a "free" charges; $\mathbf{j}$ – a conduction current;

$\mathbf{u}$ – the velocity of an ether (second terms in both (1) and (2));

$\mathbf{u}$ – the velocity of a charges (third term in (1)).

If the ether velocity **v** and the charge velocity **u** are different velocities, then we have new Hertz's equations:

$$rot(\mathbf{H}) = \frac{\partial \mathbf{D}}{\partial t} + rot[\mathbf{D}x\mathbf{v}] + \rho\mathbf{u} + \mathbf{j}; \quad (5)$$

$$rot(\mathbf{E}) = -\frac{\partial \mathbf{B}}{\partial t} + rot[\mathbf{B}x\mathbf{v}]; \quad (6)$$

$$div(\mathbf{B}) = 0; \quad (7)$$

$$div(\mathbf{D}) = \rho; \quad (8).$$

Namely the "freedom" of the ether velocity **v** does new Hertz's equations (5) – (8) very flexible for the explanation of all electromagnetic experiments.

First of all we'll receive now the equations (5) – (8) by the transformation of the differential form of the Maxwell's equations. We will also to examine the invariant character of the Hertz's equations.

Let reference frames S and S1 are moving in arbitrary manner. The coordinates of a point in the frames are linked so:

$$t = t_1$$
$$\mathbf{r} = \mathbf{r}_1 + \mathbf{R}(t) \quad (9)$$

Here $\mathbf{R}(t)$ is the vector from zero point of frame S to zero point of frame S1. It depends only from t. Then velocities in the reference frames are linked so:

$$\mathbf{u} = \mathbf{u}_1 + \mathbf{V}(t), \quad \mathbf{V}(t) = d\mathbf{R}(t)/dt \quad (10)$$

Let some vector has an image $\mathbf{f}(\mathbf{r},t)$ in S. Then it has image $\mathbf{f}^*(\mathbf{r}_1,t) = \mathbf{f}(\mathbf{r}(\mathbf{r}_1 + \mathbf{R}(t)),t)$ in S1. Then partial derivatives will linked in such manner:

$$\partial \mathbf{f} / \partial x_k = (\partial \mathbf{f}^* / \partial x_{1k})(\partial x_{1k} / \partial x_k) = \partial \mathbf{f}^* / \partial x_{1k} \quad (11)$$

$$\partial \mathbf{f} / \partial t = \partial \mathbf{f}^* / \partial t + \Sigma_k (\partial \mathbf{f}^* / \partial x_{1k})(\partial x_{1k} / \partial t) = \partial \mathbf{f}^* / \partial t - (\mathbf{V}(t)\nabla)\mathbf{f}^* =$$

$$= \partial \mathbf{f}^* / \partial t - [rot[\mathbf{f}^* \times \mathbf{V}(t)] + \mathbf{V}(t)div(\mathbf{f}^*)] \quad (12)$$

Here "$\nabla$", "rot" and "div" are derivatives with respect to $\mathbf{r}_1$.

Let now we have in S the usual motionless Maxwell's equations with convectional current. Then in S1 they will have such image:

S

$$\begin{cases} rot(\mathbf{H}) = \frac{\partial \mathbf{D}}{\partial t} + \rho\mathbf{u} + \mathbf{j}; & (13) \\ rot(\mathbf{E}) = -\frac{\partial \mathbf{B}}{\partial t}; & (14) \\ div(\mathbf{B}) = 0; & (15) \\ div(\mathbf{D}) = \rho; & (16). \end{cases} \Rightarrow$$

S1

$$\begin{cases} rot(\mathbf{H}^*) = \frac{\partial \mathbf{D}^*}{\partial t} - [rot[\mathbf{D}^* \times \mathbf{V}(t)] + \mathbf{V}(t)div(\mathbf{D}^*)] + \rho^*(\mathbf{u}_1 + \mathbf{V}(t)) + \mathbf{j}^* = \\ = \frac{\partial \mathbf{D}^*}{\partial t} - rot[\mathbf{D}^* \times \mathbf{V}(t)] + \rho^*\mathbf{u}_1 + \mathbf{j}^*; & (17) \\ rot(\mathbf{E}^*) = -[\frac{\partial \mathbf{B}^*}{\partial t} - rot[\mathbf{B}^* \times \mathbf{V}(t)]]; & (18) \\ div(\mathbf{B}^*) = 0; & (19) \\ div(\mathbf{D}^*) = \rho^*; & (20). \end{cases}$$

To eliminate the transition velocity **V(t)** from the equations in frame S1 we must to add the additional terms to equations (13) and (14) :

$$\text{rot}[\mathbf{D} \times \mathbf{w}_1] \quad \text{and} \quad -\text{rot}[\mathbf{B} \times \mathbf{w}_2] \qquad (21)$$

Here $\mathbf{w}_1$ and $\mathbf{w}_2$ are some arbitrary velocities. The terms (21) give us in S1 the same terms with the transition velocity **V(t)** as in (17) – (18) but with another signs. Then all terms with the transition velocity **V(t)** will be eliminated from (17) – (18). We will consider now that

$$\mathbf{w}_1 = \mathbf{w}_2 = \mathbf{v} \qquad (22)$$

is the ether velocity.

Then equations (13) – (16) with the additional terms (21) will have the same form as the Hertz's equations (5) – (8). Therefore the Hertz's equations (5) – (8) are invariant equations.

Moreover the Hertz's equations (5) – (8) are invariant equations not only with respect to Galilean's transformations but also with respect to arbitrary movement of the reference frames.

The "freedom" of the ether velocity **v** in Hertz's equations (5) – (8) means the necessity to built an additional theory for the ether. Such theory for the mechanical ether tried to built Stocs, for example. We do not know the physical nature of the ether. But for the Hertz's equations we need of only the velocity **v** of the ether in our laboratory frame. It is well known that all electromagnetic experiments (excluding Fizeau's experiment, see below) can be explained if the ether is motionless both in the Space (in the frame of " motionless stars") ("Lorentz's ether") and near the surface of big massive cosmic bodies (planets, stars) (in the frame which motionless with respect to the center of mass of the bodies). Near the cosmic body the ether don't rotate with the body. It's movement is only progressive movement with the center mass of the body. We don't need to know the ether velocity in the transition layer between the Space and the body, because we haven't any precision experiments in the layers.

In the picture of an ether I must to do special remark about the famous Fizeau's experiment (measurement of light velocity in a moving water). Theoretical explanation of the experiment was done by Fresnel as a case of "partial ether dragging" by moving charged bodies. But below I will show that it is simpler to explain the experiment with the help of "mean velocity" of light in dielectric medium. Therefore Fizeau's experiment hasn't any relation to the ether theory.

So we have **v** = 0 in the Hertz's equations (5) – (8) for all experiments in which we can consider that our earth laboratory frame is motionless with respect to the center mass of the Earth. Also **v** = 0 far from the Earth in the frame of "motionless stars". If our laboratory frame is moving with the velocity **w** with respect to any of these frames than **v** = - **w** in the Hertz's equations. We can experimentally discover the movement of our laboratory frame with respect to the Earth ether (see, for example, Sagnac experiment, Michelson experiment [2-4]).

Electromagnetic relativity principle in the Hertz's theory means that electromagnetic process will be the same in the reference frames in which movements of a charges and the ether are the same.

### **Invariant nature of the fields**.

Invariant nature of the electromagnetic fields in Hertz's theory means that all field vectors don't change their both values and directions in different reference frames. It not coincides with special relativity theory (SRT). But as the result we have more simple language in Hertz's theory than in SRT one.

Let us consider here two simple examples. First example is the movement in vacuum in the earth laboratory frame of point charge Q along of moving charged filament. Second example is the movement in vacuum in the earth laboratory frame of a contour with current **J** along of the same moving charged filament. Let all velocities are constants and have very small values.

First of all in Hertz's theory we must to determine the invariant Lorentz's force:

$$\mathbf{F} = Q(\mathbf{E} + [(\mathbf{u} - \mathbf{v}) \times \mathbf{B}])  \qquad (23)$$

Here: **u** – velocity of charge Q; **v** – velocity of the ether.
The force (23) is invariant because it depends only from the relative velocity. Q and field vectors are invariants. Note here that force (23) can to do the mechanical work not only in electrical field but also in magnetic field in common case. The force (23) can be modified by multiplication on $(1 - ((\mathbf{u}-\mathbf{v})/c)^2)^{1/2}$ to take into consideration an experimental results in the case of big velocities. But here we don't need in the modification.

We will to consider that velocity of the ether **v** = 0 in our Earth laboratory frame S. Our examples don't depend from t. Therefore the equations (5) – (8) will be now in the form:

$\text{rot}(\mathbf{B}) = \mu_0 q \mathbf{u}$ \qquad (24)
$\text{rot}(\mathbf{E}) = 0$ \qquad (25)
$\text{div}(\mathbf{B}) = 0$ \qquad (26)
$\varepsilon_0 \text{div}(\mathbf{E}) = q$ \qquad (26)

Here: q – density of charge on the filament; **u** – velocity of the filament.
We see that magnetic field in the frame S is created by the convectional current q**u**. On the charge Q with velocity **u** acts the force:

$$\mathbf{F} = Q(\mathbf{E} + [\mathbf{u} \times \mathbf{B}])  \qquad (27)$$

In the frame S the languages of both Hertz and SRT are the same: on the charge Q acts electric and magnetic fields.

Now let the frame S1 is connected with moving charge Q. In the frame S1 velocity of the charge Q is **u**= 0 and the ether velocity is **v** = - **u** . Then in equations (5) – (8) we can to eliminate the term q**u**. But the terms with the ether velocity give us:

-rot [-**u** x **E**] = (**E**∇)**u** – (**u**∇)**E** + **u**div(**E**) – **E**div(**u**) = q **u** \qquad (28)
-rot [-**u** x **B**] = (**B**∇)**u** – (**u**∇)**B** + **u**div(**B**) – **D**div(**u**) = 0 \qquad (29)

Here we take into account that velocity **u** don't depend from coordinate and the field gradients have the direction to the filament, i.e. they are perpendicular to the **u** .

Therefore in the frame S1 the equations (24) – (26) and the force (27) will be the same as in S. But conventional current in S1 is due to the ether movement! Therefore Hertz tells us that convectional current is due to the relative movements of charges and ether! I.e. convectional current don't depend from reference frame.

Further, in S1 the Hertz's language is different from SRT language. Namely, according to Hertz on the charge Q in S1 acts the same electric and magnetic field as they are in S. According to SRT on the charge Q in S1 acts only electric field, which is equal to the sum of the terms in round brackets in (27). I.e. Hertz tells us that magnetic field can act on a motionless charge if it is streamlined by the ether!

Now let us consider the movement of the small electroneutral contour with the current **J**

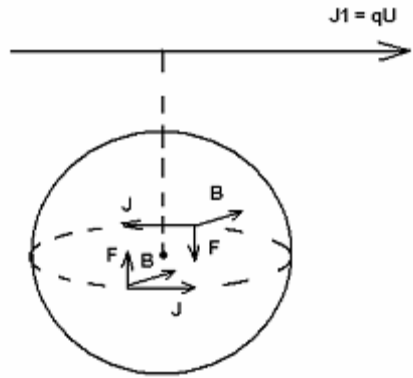

1. The small contour with current for the field B measuring.

along the same moving charged filament. It is very interesting example. The fields in the area of the contour we will consider as homogeneous because small radius of the contour. In the frame S the contour is motionless and there are turning couple of Amper's forces **F** on opposite sides of the contour as shown on the picture 1. The Amper's forces will turn the contour until both the filament and the contour will lie in the same plane. The Amper's forces let us to measure the field **B** in the area of the contour. Therefore in the frame S the contour is the device for the magnetic field measuring.

In Hertz's theory the movement of the electroneutral contour with arbitrary velocity along the filament don't influences on the couple of Amper's forces because additional Lorentz's forces on both positive and negative charges eliminate one another. I.e. the moving contour will measure the same magnetic field as the motionless contour! It is logically right because otherwise the velocity of the measuring device must be included in a magnetic field determination and in the Maxwell's equations.

Let now the frame S1 is connected with the contour and contour velocity is **u** in S . Then in S1 both the contour and the filament are motionless but contour is turning! It is because in S1 the ether is moving and magnetic field is the same as in S! I.e. we can to determine in S1 our movement with respect to the ether with the help of such experimental devices for the magnetic field measuring.

The SRT gives us, of course, the same turning of the contour in the example. But the language of the SRT is more difficult then it is in Hertz's theory. Consider it. All forces will be invariant because small velocities in the example. Let **E** and **B** are the fields in the frame S where contour is motionless. Let now the contour is moving in S with the velocity **u** . Then positive ions in the contour will have velocity **u**. The electrons will have velocities (**u+w**) and (**u-w**) on the opposite sides of the contour. Here **w** is electron velocity with respect to the contour. According to the SRT we must to find electrostatic <u>Coulomb's</u> forces, which act in the frames where the charges are motionless. Take into consideration only the linear field transformations (because of small velocities) we will find electric fields and Coulomb's forces, which act on the both positive and negative charges on opposite sides of the contour:

$$\mathbf{F}_1 = Q\mathbf{E}_1^* = Q(\mathbf{E} + [\mathbf{u} \times \mathbf{B}]) \qquad (30)$$
$$\mathbf{F}_2 = -Q\mathbf{E}_2^* = -Q(\mathbf{E} + [(\mathbf{u+w}) \times \mathbf{B}]) \qquad (31)$$
$$\mathbf{F}_3 = -Q\mathbf{E}_3^* = -Q(\mathbf{E} + [(\mathbf{u-w}) \times \mathbf{B}]) \qquad (32)$$

Now we can to find resulting Coulomb's forces on the opposite sides of the contour:

$$\mathbf{F}^*_1 = \mathbf{F}_1 + \mathbf{F}_2 = -Q([\mathbf{w} \times \mathbf{B}]) \qquad (33)$$

$$\mathbf{F}^*_2 = \mathbf{F}_1 + \mathbf{F}_3 = - Q( [-\mathbf{w} \times \mathbf{B}]) \qquad (34)$$

It is the same couple of forces as Amper's forces in the frame S! I.e. the SRT in case of small velocities gives us the same turning of the contour as it is in Hertz's theory. But the language of the SRT in that case is bad because the same device is measuring both magnetic field in frame S and electric field in frame S1! It is not very nice. The Hertz's explanation of the example is free from such difficulties.

**Experiments of Rowland H., Roentgen W., Eichenwald A., Wilson H..**

The most rigorous criticism of the original Hertz's equations (1) – (4) was excited by their contradiction to these experiments with rotating disk capacitor [2-4]. But the equations (5) – (8) very simple explain all these experiments.

To consider these experiments we will change the rotating disk capacitor with dielectric disk inside on the progressive moving parallel layers (see picture 2) [3].

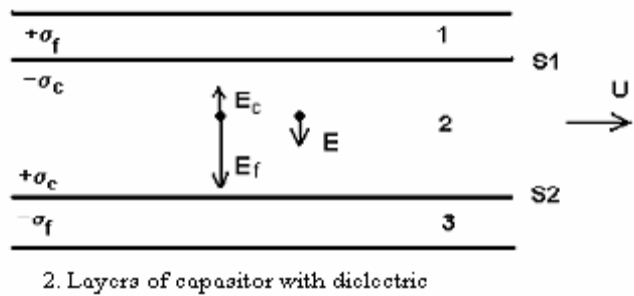

2. Layers of capacitor with dielectric

Here: 1, 3 – metallic layers; 2 – dielectric layer; S1, S2 – boundaries of the layers;
$\sigma_f$ – free charge surface density; $\sigma_c$ – bound charge surface density;
$\mathbf{E}_f$, $\mathbf{E}_c$ – electric fields of the charges;
$\mathbf{E} = ( \mathbf{E}_f + \mathbf{E}_c )$ – resulting electric field in the dielectric.

We will to consider that velocity of the ether $\mathbf{v} = 0$ in our earth laboratory frame. Our examples don't depend from t. Therefore the equations (5) – (8) will be now in the form:

$$\text{rot}(\mathbf{B}) = \mu_0(\rho_f \mathbf{u}_f + \rho_c \mathbf{u}_c) \qquad (35)$$
$$\text{rot}(\mathbf{E}) = 0 \qquad (36)$$
$$\text{div}(\mathbf{B}) = 0 \qquad (37)$$
$$\varepsilon_0 \text{div}(\mathbf{E}) = \rho_f + \rho_c \qquad (38)$$

Here all volume charge densities $\rho = 0$ in the volumes but $\rho = \infty$ on the boundaries. The surface densities of the charges we will mark as $\sigma$.

We will consider the fields only near one bounder S1, for example (the full measured magnetic fields are in factor 2 greater, of course). From (35) and (38) we obtain follow boundary conditions:

$$2B = \mu_0 | \sigma_f \mathbf{u}_f - \sigma_c \mathbf{u}_c | \qquad (39)$$
$$\varepsilon_0 E = \sigma_f - \sigma_c , \text{ here: } \varepsilon_0 E_f = \sigma_f , \varepsilon_0 E_c = \sigma_c \qquad (40)$$

Now we must to postulate the usual "matter" equation. We write it as the equation for the Lorentz's forces, which act inside the dielectric on bound charges:

$$\mathbf{F}_f = \varepsilon \mathbf{F}^* = \varepsilon (\mathbf{F}_f + \mathbf{F}_c + \mathbf{F}_{ext}) = \varepsilon \{(\mathbf{E}_f + \mathbf{E}_c + \mathbf{E}_{ext}) + [ \mathbf{u}_c \times \mathbf{B}_f ] + [ \mathbf{u}_c \times \mathbf{B}_c ] + [ \mathbf{u}_c \times \mathbf{B}_{ext} ] \} \qquad (41)$$

Here: $\varepsilon$ - permittivity of the dielectric; "ext" – external fields.
From (41) we can find:

$\mathbf{E}_c = (1- \varepsilon)( \mathbf{E} + [\mathbf{u}_c \times \mathbf{B}])$ or

$\mathbf{E}_c = \{ ((1- \varepsilon)/\varepsilon)( \mathbf{E}_f + \mathbf{E}_{ext} + [\mathbf{u}_c \times \mathbf{B}_f ] + [ \mathbf{u}_c \times \mathbf{B}_{ext} ]) - [\mathbf{u}_c \times \mathbf{B}_c ] \}$ (42)

Note that external magnetic field exists only in Wilson's experiment (see case d) below). In other cases we haven't any external fields. Let the dielectric is not magnetic. Then we can eliminate the terms with $\mathbf{B}_f$, $\mathbf{B}_c$ from (41),(42) because they are very small.

Now we can to explain all the experiments:
a) only metallic disk is moving with $\mathbf{u}_f$. Then we have measured magnetic field:

$B = 0.5\mu_0\sigma_f u_f = 0.5\mu_0\varepsilon_0 E_f u_f = 0.5\mu_0\varepsilon_0\varepsilon E u_f$ (43)

It was confirmed by the experiment.

b) only dielectric is moving with $\mathbf{u}_c$. Then we have measured magnetic field:

$B = 0.5\mu_0\sigma_c u_c = 0.5\mu_0\varepsilon_0 E_c u_c = 0.5\mu_0\varepsilon_0(\varepsilon - 1)E u_c$ (44)

It was confirmed by the experiment.

c) both metallic disk and dielectric disk are moving with $\mathbf{u}$. Then we have measured magnetic field:

$B = 0.5\mu_0(\sigma_f - \sigma_c)u = 0.5\mu_0\varepsilon_0 E u$ (45)

It was confirmed by the experiment.

d) only dielectric is moving with $\mathbf{u}_c$ in the external field $\mathbf{B}_{ext}$. Then we have measured electric field:

$E = E_c = ((\varepsilon - 1)/\varepsilon)| [ \mathbf{u}_c \times \mathbf{B}_{ext} ] | = ((\varepsilon - 1)/\varepsilon)u_c B_{ext}$ (46)

It also was confirmed by the experiment.

Therefore the Hertz's equations (5) – (8) very simple explain all these classical experiments. Moreover, the "matter" equation (41) is as important for the explanation as the equations (5) – (8).

**Fizeau's experiment and some other experiments.**

In the famous Fizeau's experiment was measured the light velocity in a moving water. It was shown that the velocity is :

$V = C/n – [1 – 1/n^2]U$ (47)

Here : C – light velocity in vacuum; $n = \varepsilon^{1/2}$ – refraction coefficient ; $\varepsilon$ - dielectric permittivity of the matter ; U – the matter velocity.

In the usual criticism Hertz's equations (1) – (4) contradict to the Fresnel's formula (47), because they contain "full ether dragging" while (47) contain only "partially ether dragging".

I will show below that the Fresnel's formula (47) is not connected with any form of the ether! I.e. the Hertz's equations (5) – (8) don't contradict to the formula.

I will show that (47) is simply formula for the mean velocity of the light in a matter, because the light movement in a matter is discrete process!

In quantum mechanics the process of light movement in a matter is a sequence of an absorptions and reemissions of a light quantum by molecules or atoms of the matter.

Let L is the mean free path of the light quantum between two molecules or two atoms of a matter and $\tau$ is the mean time of life of the molecule or atom in the excited state. During the time $\tau$ the light quantum is moving together with the atom with the matter velocity. Let $\mathbf{U} = \mathbf{W} - \mathbf{v}$ is the matter velocity with respect to the ether. In our earth laboratory frame the ether velocity $\mathbf{v} = 0$. Let the matter is moving to the light source. Than the mean velocity of the light quantum is :

$$V = (L - UT) / T = L / T - U \qquad (48)$$

Here T is the full time of the movement of the quantum between two atoms :

$$T = \tau + t \qquad (49)$$

and t is time of free path of the quantum. We can find t from :

$$tC + tU = L$$

i.e.:

$$t = L / (C + U) \qquad (50)$$

Then :

$$T = \tau + L / (C + U) \qquad (51)$$

Then we have from (48) and (51) mean velocity of the quantum:

$$V = L / [\tau + L / (C + U)] - U$$

or :

$$V = [L/C - \tau (1+k)k] / [\tau (1+k)/C + L/C^2 ] \; ; \; k = U/C \qquad (52)$$

Let as usually $k \ll 1$. Than Tailor's expansion of the (52) until the first order in k is:

$$V = C/(1+\tau C/L) - [1 - 1/(1+\tau C/L)^2]U \qquad (53)$$

The formula (53) is the Fresnel's formula (47) if we determine n as:

$$n = (1 + \tau C/L) \qquad (54)$$

The Hertz's equations (5) – (8) don't contain ether velocity for the experiment ($\mathbf{v}=0$), because laboratory system is motionless with respect to the center mass of the Earth. Therefore the Hertz's equations is the same as Maxwell's equations for the motionless matter. The equations give us the electromagnetic wave velocity C/n in the matter. Therefore we can to get right velocity (47) for the moving matter by two ways. On the first way we must to do some additional artificial transformation of both time and space intervals to do continuous wave velocity equal to the mean quantum velocity (47). The way is realized in Lorentz's transformations. On the second way we must to include the matter velocity U in the "matter" equation. I will not to do that for a common case. But for our example we can simply to determine the new refraction coefficient $n^*$:

$$C/n - [1 - 1/n^2]U = C / n^*$$

Then:

$$n^* = n \{C / [C - (1 - 1/n^2)Un] \} \qquad (55)$$

Now we will briefly mention here a few other important experiments because their explanations are very simple on the base of the Hertz's equations (5) – (8).

The well known experiment of Michelson A. and Morley E., and the experiment of Trouton F. have the negative results because their devices were motionless with respect to motionless ether near the Earth surface [2-4].

The important experiments of Michelson and Sagnac showed us the movement of the laboratory frames with respect to the earth ether [2-4].

**Relativistic Doppler effect for Hertz electrodynamics.**

To add the relativistic Doppler effect to the Hertz electrodynamics we must to consider that the ether act on the electron energy levels, when the atom is moving with respect to the ether.

A long before the special relativity theory (SRT) was established the formula (56) by both the experimental work of Kaufmann and theoretical works of Abraham, Lorentz, Poincare:

$$m = \frac{m_0}{\sqrt{1-\left(\frac{u}{c}\right)^2}} = \frac{m_0}{\sqrt{1-(\beta)^2}} \quad (56)$$

Here: $m_0$ – is electron mass of motionless electron ;
 u – is electron velocity;
 c – is light velocity.

The laboratory system in the experiments was motionless with respect to the Earth. Consequently we can to consider the laboratory system as motionless with respect to the ether, if we will to neglect in the experiments by the Earth rotation around it's axis. Then we can to consider "u" as the electron velocity with respect to the ether.

We can to consider the formula (56) as a result of the ether action on the moving electron.

The Newton law for the electron movement is:

$$m\mathbf{a} = \mathbf{F}_0 \quad (57)$$

Where: a – is the acceleration of the electron;
 $F_0$ – is the force on the motionless electron.

Let to write the equation (57) in such form:

$$m_0\mathbf{a} = \sqrt{1-(\beta)^2}\,\mathbf{F}_0 = \mathbf{F} \quad (58)$$

I.e. we will consider that we have constant mass of the electron and different resulting force. It means the existence of the ether force:

$$\mathbf{F}_{eth} = (\sqrt{1-(\beta)^2} - 1)\mathbf{F}_0 \quad (59)$$

And the resulting force is:

$$\mathbf{F} = \mathbf{F}_0 + \mathbf{F}_{eth} \quad (60)$$

Let the atom nuclear is motionless with respect to the ether. Then the electron movement with respect to the both nuclear and ether is determined by a force $F_0$. The force is doing the work:

$$dA_0 = (\mathbf{F}_0 d\mathbf{r}) \quad (61)$$

Where: dr - is the shift in the electron position.

Let $A_0$ or it's part determine the electron energy level $E_0$ for the motionless atom. The rules for the energy levels calculations are described by the quantum mechanics.

If the atom is moving with respect to the ether we must to add an additional ether force on the electron. Then we can to write the work of the resulting force as:

$$dA = (\mathbf{F}d\mathbf{r}) = \sqrt{1-(\beta)^2}\,(\mathbf{F}_0 d\mathbf{r}) = \sqrt{1-(\beta)^2}\,dA_0 \qquad (62)$$

Accordingly we can to write the new electron levels as:

$$E = \sqrt{1-(\beta)^2}\,E_0 \qquad (63)$$

Let us to consider the laws of both the energy conservation and the linear momentum conservation with respect to the quantum emission process.

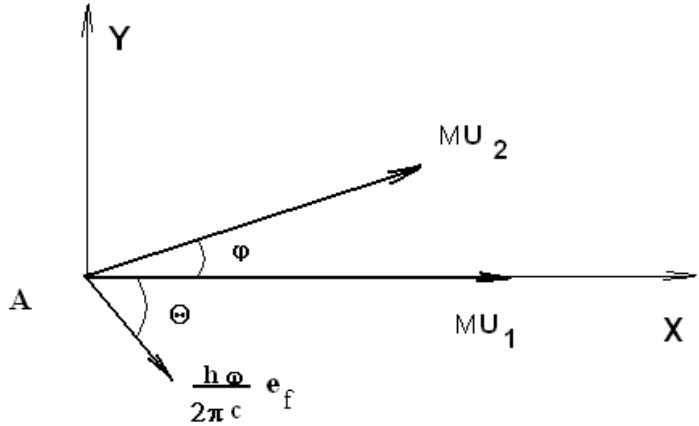

Fig.3. Linear momentums of both the atom and the quantum in the emission process.

On the Fig.3: $U_1, U_2$ – are the atom velocities before and after the emission;
 M – is the atom mass;
 h – is the Plank constant;
 ω – is the cyclic frequency;
 $\mathbf{e}_f$ – is the unit vector of the emission direction.

In the emission process the electron goes from the energy level $E_1$ to the energy level $E_2$.
The law of the linear momentum conservation in the closed system atom + quantum is:

$$Mu_1 = Mu_2 \cos\varphi + \frac{\hbar\omega}{c}\cos\vartheta \qquad (64)$$

$$0 = Mu_2 \sin\varphi - \frac{\hbar\omega}{c}\sin\vartheta \qquad (65)$$

The equation (65) let us to link the angles φ and θ. But we will consider that $Mu_2 \gg \hbar\omega/2\pi c$. Then from (65) we will receive sin φ = 0 (or cos φ = 1). Then (64) give us:

$$M(u_1 - u_2) = \frac{\hbar\omega}{c}\cos\vartheta \qquad (66)$$

The law of the energy conservation:

$$\frac{Mu_1^2}{2} + E_1 = \frac{Mu_2^2}{2} + E_2 + \hbar\omega \qquad (67)$$

or

$$\frac{1}{2}M(u_1 - u_2)(u_1 + u_2) + (E_1 - E_2) - \hbar\omega = 0 \qquad (68)$$

With (66) we can to write (68) as:

$$\frac{\hbar\omega}{c}\cos\vartheta \frac{u_1 + u_2}{2} + \left(\sqrt{1-(\beta_1)^2}E_{10} - \sqrt{1-(\beta_2)^2}E_{20}\right) - \hbar\omega = 0 \qquad (69)$$

If $u_1 \approx u_2 \equiv u$ and $E_{10} - E_{10} = \hbar\omega_0/2\pi$, then we can write (69) as:

$$\frac{\hbar\omega}{c}u\cos\vartheta + \sqrt{1-(\beta)^2}\hbar\omega_0 - \hbar\omega = 0 \qquad (70)$$

or

$$\omega = \frac{\omega_0\sqrt{1-(\beta)^2}}{1 - \beta\cos\vartheta} \qquad (71)$$

The formula (71) has the same form as the relativistic Doppler effect in the SRT. But the means of the velocity U are different, in common case, in the SRT and in the ether theory.

The absorption process of the quantum by the atom – receiver, which is moving in the ether, is considered by the same formulas. Only some terms in (66) and (67) will to have other signs. But final result will be the same formula (71). It gives us the frequency, which is necessary for the absorption of the quantum by the moving atom.

**Conclusions**.

a) Hertz's equations (5) – (8) are very flexible and can to explain all electromagnetic experiments with moving bodies without the SRT. They are invariant with respect to arbitrary moving frames.

b) Hertz's equations (5) – (8) give us the link between electrodynamics and the massive cosmic bodies movements, because we can speak not about the ether velocity **v** but about the massive cosmic body velocity **v** in our laboratory frame.

c) The capacity of the Hertz's equations (5) – (8) for work means that the expansion of the SRT out of the classical mechanics and classical electrodynamics is not absolute necessary direction for the physical theory development.

d) Lorentz's transformations can be used only if our theoretical model of an electromagnetic phenomenon is not complete model. They can be used also for the mathematical reformulation of an electromagnetic tasks with the aim to do the decision of the tasks more easily as it did Lorentz himself in he's proceedings.

**Literature.**